*Research Article*

# Exploring Preferences for Transportation Modes in the City of Munich after the Recent Incorporation of Ride-Hailing Companies

**Maged Shoman[1] and Ana Tsui Moreno[1]**

## Abstract

The growth of ride-hailing (RH) companies over the past few years has affected urban mobility in numerous ways. Despite widespread claims about the benefits of such services, limited research has been conducted on the topic. This paper assesses the willingness of Munich transportation users to pay for RH services. Realizing the difficulty of obtaining data directly from RH companies, a stated preference survey was designed. The dataset includes responses from 500 commuters. Sociodemographic attributes, current travel behavior and transportation mode preference in an 8 km trip scenario using RH service and its similar modes (auto and transit), were collected. A multinomial logit model was used to estimate the time and cost coefficients for using RH services across income groups, which was then used to estimate the value of time (VOT) for RH. The model results indicate RH services' popularity among those aged 18–39, larger households and households with fewer autos. Higher income groups are also willing to pay more for using RH services. To examine the impact of RH services on modal split in the city of Munich, we incorporated RH as a new mode into an existing nested logit mode choice model using an incremental logit. Travel time, travel cost and VOT were used as measures for the choice commuters make when choosing between RH and its closest mode, metro. A total of 20 scenarios were evaluated at four different congestion levels and four price levels to reflect the demand in response to acceptable costs and time tradeoffs.

The past few years have witnessed a significant growth of gig-companies, known as transportation network companies (TNCs), operating on-demand and app-based, prearranged services, which are also referred to as ride-hailing (RH), ride-sourcing and ride-matching (*1*). The original term used for such services was the new online enabled transportation service (NOETS). RH services are offered through an online platform that connects customers demanding a ride to drivers offering a ride supply using their own vehicles. The GPS capability of the online platform allows the driver to determine the passenger's pickup location and keeps the passenger updated about the driver's location and arrival time.

In Germany, Uber began its operations during early 2013 in Berlin, the capital city. Berlin's taxi companies put up a ferocious resistance, including legal action. Meanwhile, Uber kept proceeding with its plans to expand into operations in Hamburg, Cologne, Stuttgart and Dusseldorf (*2*). However, after several claims and legal suits filed against Uber citing them as an unfair market player, and after the service being banned in several cities, Uber representatives gave up on the German market. Berlin and Munich are its only operational grounds now, with limited services, enabling it to operate while complying with the German laws. Besides Uber, other ride-sharing services such as Cabify and BlaBlacar operate in Munich and most German cities today.

Previous studies on RH services reflect their impact on urban mobility through measures such as modal shares. A literature review of the available work reflected the struggle to acquire data from RH companies about their vehicles, drivers and passengers. Several authors tried different approaches to acquire adequate and reasonable RH data using surveys and interviews. An innovative approach to dissect the market was adopted by (*3*), in which one of the authors personally drove for Uber and Lyft. Studies in major American cities found that RH companies are responsible for a 6% reduction in transit

[1]Department of Civil, Geo and Environmental Engineering, Technical University of Munich, Munich, Germany

**Corresponding Author:**
Maged Shoman, magedshoman@gmail.com

use (*4*). In the same work the authors found that if Uber or Lyft services did not exist, around 55% of ride-hailing trips would not take place or would be made by walking, biking or transit. Another survey found 21% using RH for commuting but a larger proportion using transit for that purpose, with RH popularity higher in late evenings and night and lower in the morning and evening rush (*5*). (*1*) found that if RH did not exist in San Francisco, 33% of its users would switch to public transit.

Despite the consensus that RH services are more efficient than their fellow modes, literature on the topic is quite limited. The available research stresses the importance of data to study the impacts of such rapidly growing services. To fill the gap in academic literature and aid in studying the impacts of RH services, we designed a survey to understand how travelers in the city of Munich value their time when using RH. Studying the influence of RH using the value of time (VOT) addresses willingness to pay by different income groups, which was a limitation of several published pieces of research.

The objective of this paper is to estimate the impact of RH services on other transportation modes based on how users respond to changes in auto, metro and RH travel time and cost in an online survey.

The paper is structured as follows. A literature review presents the research related to acceptance of and willingness to use RH in different countries and the impacts of RH services on other transportation modes. This is followed by a section on methodology, covering the design of the survey and how its results are incorporated into the mode choice of an existing travel demand model. Analysis and results are added after the methodology. Finally, conclusions are presented in the last section.

## Literature Review

Once RH companies were accepted into the market and demand for such services started increasing, studies revealed that the RH users mostly use it for social trips and other similar purposes, but rarely use it for work purposes, compared with transit (*5*). In New York, RH services are commonly used for activities such as social activities, shopping, entertainment, and so forth (*6*). In other studies, RH has satisfied commuter demands for rapid response during unpredictable weather conditions, using surge pricing to increase its supply (*7*). (*4*) found that RH services provide an unprecedented level of convenience and the survey responses showed that 37% would choose RH because of the struggle to find parking, and 33% would do so to avoid drunk driving.

As RH companies gained popularity, their services grew very fast in a short time. A 2019 paper showed that Uber had generated 31 million trips and 52 million passengers since 2013 (*8*). It was not long until RH companies affected taxi ridership in ways that harmed the taxi industry, with taxi ridership decreasing in most of the studies, specifically by 65% in San Francisco from 2012 to 2014 (*8*), an 8% decrease in taxi rides per hour in New York (*7*), an 18% drop in passenger trip numbers per day in Beijing from 2012 to 2015 (*9*), an 18% drop in Toronto (*10*) and a 15% decrease in Dubai after RH companies' entrance into the market (*8*). When comparing RH vehicles to traditional taxis, research by (*11*) in five cities in the US found that RH vehicles have a higher efficiency rate in relation to trips made compared with taxis. (*12*) found that the higher capacity of RH vehicles leads to a significant decrease in congestion and exhaust emissions. The effect of RH's entrance into the market on other modes of transport is noteworthy. The analysis of the changes in mode shares after the introduction of RH services in Toronto from 2011 to 2016 shows that RH changed from 0% to 24.1%, taxi from 22.8% to 5.2%, transit from 16.3% to 20.3%, auto from 44.6% to 21.4% and active modes from 16.3% to 29.1% (*10*). In the same study, RH (Uber) was found to cost less than using a taxi (taxi fare: $7.20 for a 6km trip) but more expensive than transit (transit fare: $3.25).

Seeking to determine RH users' market segment, (*10*) found that in Toronto, people mostly using TNCs are between 20 and 39 years old with only 2% of users aged 60 or over. The majority of the trips took place between late night and 5 a.m. The same study found that households with higher earnings use RH (54%) more than do low income groups (2.6%). (*13*) also found similar results for the market segment demand, which is explained by the common use of technology among younger age groups. A recent study using revealed preference (RP) surveys to model the demand of users using rewards incentives found that energy savings are valued more highly than cost savings and that the acceptance of reward system use is more common among lower income segments (*14*). Money and convenience were considered more important by RH users in China than risks such as privacy and security of using the service (*15*).

To investigate the main determinants of riders' choices, (*16*) used a stated preference (SP) survey to conduct a feasibility study of an interurban RH service on a commuter sample in Salerno metropolitan area. (*17*) also used an SP survey on commuters in Chicago to determine the influence of trip parameters such as time and cost on demand. The use of SP surveys is common in studies with different transportation modes and was utilized by (*18*) to study the influence of level-of-service on the use of autonomous and RH vehicles.

Key findings from a report published by the Transportation Research Board in 2018 focused on understanding the interaction between different transportation modes, mainly RH services in five regions

(Chicago, Los Angeles, Nashville, Seattle, and Washington D.C.) (*19*):

- Highest use of RH occurs during late hours of the day and weekends.
- Top concerns for users who would shift from transit to RH were transit travel and waiting times.
- RH is used by all income groups.
- The use of RH is affected by the decrease in auto ownership with frequent users reporting no autos per household.

Uber had a complementary effect in transit cities with low transit ridership and a substitution effect in cities with high transit. This is because of Uber's ability to provide additional flexibility when transit supply was insufficient (*20*). The effects of RH services have been studied from different viewpoints with simulation results presenting potentials, but also limitations, of such services (*21*, *22*), with the main issue being efficiently supplying the service on demand at the right time. Many RH studies also had limitations with respect to estimating the willingness of commuters to share a ride.

## Methodology

### SP Online Survey

The SP survey was designed to be carried out through an online survey. The purpose of the survey was to determine travelers' preferred mode for a given trip, varying mode-specific attributes and the trip purpose. Auto, transit and RH were the alternatives in the choice set. A 20-question survey was created in Limesurvey (limesurvey.org). The survey was designed in four parts.

(1) Sociodemographic profile. Individuals provided their gender, age, occupation, residence area type, residence period in Munich, distance to nearest transit stop, auto license ownership, and the number of people, workers, children and autos in a household.
(2) Used mode of transport on an average day. Individuals were asked about the most common transportation mode used for different trip purposes.
(3) Willingness to use RH services for HBW (home-based work) trips. Individuals stated their preference from a choice set of three transportation modes (RH, auto and transit) across three scenarios varying cost and time.
(4) Willingness to use RH services for HBO (home-based other) trips. Individuals stated their preference from a choice set of three transportation modes (RH, auto and transit) across three scenarios varying cost and time. Cost and time values were exactly like the HBW scenarios.

The key objective of the scenarios presented was to acquire a confident measure of the demand for RH and the respondent's VOT in comparison with mutually exclusive alternatives—private auto and public transport. For the design of the base scenario, we used an 8 km trip from Gundermannstraße-15b to Augustenstraße-118 in Munich city. Since the most common form of RH services in Munich is Uber, UberApp (uber.com) was used to estimate the total travel time and cost. The same data was adopted from Google Maps for auto and Münchner Verkehrs- und Tarifverbund (MVV) for public transport. Variables added for consideration included walking time to vehicle, waiting time, in-vehicle time, parking search time, walking time to destination, travel cost and parking cost. The only distiction between the presented scenarios is in using the RH service for a shared ride. To consider the changes in a shared ride, total travel time was increased by 10% and 20% and the cost was reduced by half across the two additional scenarios.

The average response time to the survey was around 5 min. The survey was offered for two weeks from March 11 to 25, 2019, in both German and English, available online to travelers in the city of Munich. The survey link was emailed to several students at different universities, employees at different companies and Facebook groups. Additionally, flyers were printed in German and English with the survey barcode for easier access via smartphones and were distributed around the city center.

The raw total sample of responses was 800. However, not all respondent completed the survey or provided their sociodemographic attributes. The final sample was reduced to 500. Around 90% of the respondents took the survey in English, and respondents' gender was evenly distributed. Of English respondents, 90% were under 39 years old and only 37% owned more than one auto. Conversely, 82% of the German respondents were younger than 39 years old but 70% owned more than one auto.

To provide realistic results applicable to travelers in Munich, the survey responses were weighted to match the census data for the city of Munich. Iterative proportional fitting (IPF) was used. IPF is the most widely used mature deterministic method of allocating individuals by calculating a series of noninteger weights that reflect how representative each individual is of each constraint (*23*). Specifically, age, gender and household size distributions were used as control totals. The extension mlogit in *R* was used (*24*).

### Incorporating the Results in a Travel Demand Model

To examine the impact of travelers' willingness to use RH services, we incorporated the results of the survey into the mode choice model of the travel demand model MITO (Microscopic Transportation Orchestrator) (*25*). In this section, we will describe MITO and how its mode choice model was modified to incorporate a new transport mode: RH.

*MITO Travel Demand Model.* MITO uses microsimulation to simulate each household and person individually. It follows a disaggregated four step model with the constraint of travel time budget.

A synthetic population is generated using iterative proportional updating (IPU) at three geographical levels: county, municipality and borough (*26*). Households and persons are used as input to MITO. First, the number of trips is generated for each household. Trip generation uses sample enumeration to select the number of trips made for every household type for each purpose. Household types were defined inductively by testing 67 million possible household type definitions for each purpose (*27*). Second, a destination for every trip is assigned. The destination for mandatory trips, such as work and education trips, is already defined by the synthetic population. The destination of discretionary trips depends on the remaining travel time budget for the household. Based on Zahavi's theory of constant travel time budgets (*6*), longer trips to work will lead to shorter discretionary trips. The travel time budget is not a hard constraint for an individual household, but is rather used to influence the probabilities of choosing different destinations so that travel time remains constant over time.

The mode choice model is built as a nested logit model. It includes an auto nest (auto driver and auto passenger), a transit nest (bus, metro, train), bicycle and walking. RH was not included in the choice set because it had not yet been introduced into the German market at the time of the National Household Travel Survey (MiD2008, mobilitaet-in-deutschland.de).

The time-of-day choice model selects a preferred arrival time for trips and calculates the trip departure time based on the expected travel time by the selected mode. The trip assignment is simulated in the multiagent travel-based model, MATSim (*28*).

*Mode Choice Model with RH.* The mode choice of MITO did not include RH in the choice set. To provide a new mode choice model with RH, there were two approaches: (1) estimate a new model based on the responses to the SP online survey and calibrate the alternative-specific constants to the modal split resulting from the responses, or (2) extend the current model using an incremental logit approach (*29*), which is able to estimate the modal shares of RH based on the current modal shares and the changes in the service characteristics with respect to the existing set of alternatives. While the first option would provide more accurate results for RH and the modes included in the online survey, we would be losing the richness of the German household travel survey's large sample size (176,000 versus 500 trips), modes of transport covered (7 modes versus 3 modes) and range of trips included (across the region versus one trip in the city; six trip purposes versus two trip purposes; range of distance, etc.), as well as the nested structure of the MITO model. Alternatively, the incremental logit approach tries to reflect how users react differently to changes in a reference mode and RH travel time and cost, assuming that there are no changes in the utility of other variables or the alternative-specific constant. After careful consideration, we opted to apply an incremental logit model to incorporate the new service into the calibrated nested logit model of MITO.

The application of the model requires the econometric estimation results of a reference mode and the difference between the new mode and the reference mode. Figure 1 presents the nested model structure with the current modes in blue and the new mode to be incorporated in orange.

The structure includes RH in the transit nest since the majority of auto/ride-sharing service users would substitute such a service for public transport or auto as found by (*4*). It is argued that the nature of RH is as a public transport mode in the mobility system. Like the metro mode, RH is designed to travel faster than buses and provides a point-to-point travel option.

To define the changes in utilities for RH, we use the results from the SP survey. We assume that travel time, travel cost and VOT are the only variables that are different for the travelers when they make the choice between metro and RH. All other sociodemographic attributes, area attributes and the alternative-specific constant

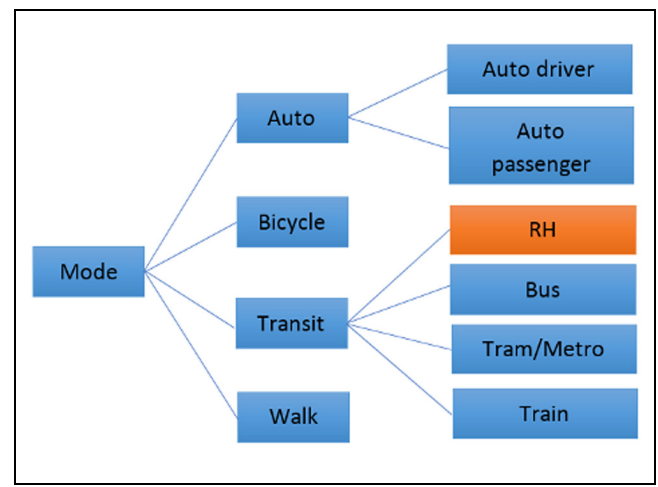

**Figure 1.** Nested model structure (current modes in blue, new ride-hailing mode [RH] in orange).

remain unchanged for the new mode, following the incremental logit model approach.

To incorporate the different sensitivity of RH users to travel time by transit, we modify the RH time coefficient by a factor equal to the ratio between the travel time coefficient of RH and that of transit. The VOT for RH users is also incorporated to calculate the generalized cost of this mode.

The utility of RH is defined by

$$U'_{\text{RH}} = U_{\text{metro}} + \beta_{\text{gCmetro}} \bullet (\text{gC}_{\text{RH}} - \text{gC}_{\text{metro}}) \quad (1)$$

$$\text{gC}_{\text{RH}} = \frac{\delta_{\text{timeRH}}}{\delta_{\text{timeMetro}}} \bullet \text{time}_{\text{RH}} + \frac{\text{distance}_{\text{RH}} \bullet \text{farePerKm}_{\text{RH}}}{\text{VOT}_{\text{RH}}} \quad (2)$$

where

$U'_{\text{RH}}$ is the utility of RH;

$U_{\text{metro}}$ is the utility of metro (reference mode);

$\beta_{\text{gCmetro}}$ is the coefficient for generalized cost of metro;

$\text{gC}_{\text{RH}}$ is the generalized cost of RH;

$\text{gC}_{\text{metro}}$ is the generalized cost of metro;

$\delta_{\text{timeRH}}$ is the coefficient of time for RH from the SP survey;

$\delta_{\text{timeMetro}}$ is the coefficient of time for transit from the SP survey;

$\text{time}_{\text{RH}}$ is the total travel time by RH;

$\text{distance}_{\text{RH}}$ is the total distance by RH;

$\text{farePerKm}_{\text{RH}}$ is the fare per km by RH; and

$\text{VOT}_{\text{RH}}$ is the value of time for RH.

After obtaining the utility for RH, the probability of selecting RH is defined by

$$P'_{\text{RH}} = \frac{\exp\left(\frac{U'_{\text{RH}} - U_{\text{metro}}}{\text{nc}}\right)}{\exp\left(\frac{(U_{\text{bus}} - U_{\text{metro}})}{\text{nc}}\right) + \exp\left(\frac{(U_{\text{train}} - U_{\text{metro}})}{\text{nc}}\right) + 1} \times P'_{\text{Transit}} \quad (3)$$

$$P'_{\text{Transit}} = \frac{\frac{\exp(U'_{\text{Transit}})}{\sum_M \exp(U_M)}}{\frac{\exp(U'_{\text{Transit}})}{\sum_M \exp(U_M)} + [1 - P_{\text{Transit}}]} \quad (4)$$

where

$P'_{\text{RH}}$ is the probability of selecting RH;

$U_{\text{mode}}$ are the utilities by transit mode (metro, bus, train);

nc is the nesting coefficient;

$P'_{\text{transit}}$ is the probability of selecting transit modes; and

$U_M$ are the utilities by mode (walk, bicycle, auto nest).

## Application and Sensitivity Analysis

The study area is the Munich metropolitan region. It is located in southern Germany and includes five core cities: Munich, Augsburg, Ingolstadt, Landshut and Rosenheim and their suburbs. The study area is delimited to municipalities that have a commuter flow higher than 25% to any of the core cities (30). Travel demand was generated for all the Munich metropolitan region; however, the RH service area only included the city of Munich city. This assumption was in line with current RH services in the region.

A total of 20 RH scenarios were simulated with different RH travel times and RH costs. Further, we also simulated one scenario without RH to compare how modal splits vary after the introduction of RH. RH travel time was assumed to equal the travel time by auto in the base scenario. We generated scenarios in which, as congestion worsens, travel time increases proportionally by a factor. Further, waiting times also increase and were added to the in-vehicle travel time to obtain the total travel time. For travel costs, we used the base fare of €1.5/km, equal to the fare that Uber has in the city of Munich. We assumed a reduction of the base fare to €0.75/km when carpooling services were offered. Also, we increased the base fare to €3.0/km and €6.0/km to consider surge pricing. The RH service characteristics varied as follows:

- Travel time (four levels): 1.0 (base); 1.1; 1.2; and 1.5 times the base travel time. They were associated with waiting times of 0, 4, 8, and 18 min, respectively.
- Travel cost (four levels): 0.75; 1.50 (base); 3.00; 6.00 in €/km.

Only RH service characteristics varied across scenarios. Therefore, we maintained the same service characteristics for transit and auto.

## Analysis and Results

### SP Online Survey

*Preliminary Analysis.* The first step was to check whether the responses from the survey should be weighted to match the general population characteristics in Munich. The most recent census data available for Germany is from 2011 (www.zensus2011.de). As shown in Table 1, the demographics of the survey respondents were different than those observed in the census data: persons between 25 and 39 years old were overrepresented, while persons older than 50 years old were highly underrepresented. To balance the sample, we applied IPF to gender, age and household size. The correlation after IPF was 0.998. Table 1 presents the share of each group in the census data and in the unweighted and weighted survey, together with the weights. Higher weights were given to groups that were underrepresented, such as respondents older than 50 (with

**Table 1.** Weighted Survey and Population Statistics

| Variable | Census data | Group share | | |
| | | Unweighted survey | Weighted survey | Weights |
|---|---|---|---|---|
| Gender | | | | |
| Male | 48.92% | 45.00% | 49.00% | 1.09 |
| Female | 51.08% | 55.00% | 51.00% | 0.93 |
| Age | | | | |
| <18 | 0.00% | 0.00% | 0.00% | 0.00 |
| 18–24 | 12.87% | 19.52% | 14.23% | 0.73 |
| 25–29 | 7.88% | 30.88% | 9.09% | 0.29 |
| 30–39 | 15.99% | 37.45% | 16.56% | 0.44 |
| 40–49 | 20.08% | 8.17% | 18.68% | 2.29 |
| >50 | 43.18% | 3.98% | 41.44% | 10.41 |
| Household size | | | | |
| 1 | 39.12% | 33.80% | 39.52% | 1.17 |
| 2 | 30.86% | 34.80% | 31.22% | 0.90 |
| 3 | 13.75% | 14.20% | 13.36% | 0.94 |
| 4 or more | 16.27% | 17.20% | 15.90% | 0.92 |

**Table 2.** Survey Modal Split across Scenarios, by Trip Purpose

| Scenario number | Trip purpose | RH | Auto | Public transport |
|---|---|---|---|---|
| Scenario 1 | HBW | 1.28 | 18.57 | 80.15 |
| (TT, C) | HBO | 2.09 | 40.33 | 57.58 |
| Scenario 2 | HBW | 4.66 | 19.19 | 76.15 |
| (1.1TT, 0.5C) | HBO | 8.04 | 36.73 | 55.23 |
| Scenario 3 | HBW | 3.84 | 24.67 | 71.49 |
| (1.2TT, 0.5C) | HBO | 6.39 | 38.28 | 55.33 |

*Note:* C = cost ; HBW = home-based work; HBO = home-based other; RH = ride-hailing; TT = travel time.

the maximum weight of 10.41), while smaller weights were given to groups that were overrepresented, such as persons between 25 and 29 (with the minimum weight of 0.29). Introducing weighted survey data could result in reduced accuracy of the models. Furthermore, the up-weightings for persons older than 50 were significant: they exaggerated by 10 times the collected responses. However, introducing the weights reduces the bias in the estimation results caused by overrepresented groups. To better control for the impact of weighting our survey responses, we estimated both the unweighted and the weighted models and compared the results. Survey respondents who selected options outside of the variable categories presented in Table 1 were very few and therefore were excluded from the study.

Table 2 presents the mode share distribution across the three survey scenarios based on the survey responses. As the scenarios are the same in both sets, RH seems to have had a higher preference when traveling on trips with a purpose other (HBO) than work (HBW). Scenario 2, with the shorter time and least cost, had the highest modal share in both sets, with 8.04% for HBO purposes and 4.66% for HBW purposes.

*Mode Choice Model Estimation.* A multinomial logit model (MNL) was used to estimate a mode choice model from the survey responses. The gmnl package in RStudio (*19*) was used to build the two models for each purpose: HBW and HBO. The wide format dataset included individual and transport characteristics as designed in the survey, removing the highly correlated variables. Weighted and unweighted responses were used to control for the impact of the weighting procedure on the sensitivity of the model. The signs of the coefficients for weighted and unweighted data remain equal (e.g., persons in households with an auto are less likely to use RH), although the magnitudes differ. Given that the weights were introduced to minimize the bias of the survey respondents in age and household size distributions, it was likely that the greater differences were in those variables. The unweighted coefficients were larger for young adults because they were overrepresented in the sample and their responses were disproportionally affecting the results compared with the population. After weighting the responses, the coefficient for young adults was reduced, correcting for this imbalance. Looking at the two sets of cross-tabulations (weighted and unweighted), the estimates still look reasonable and the differences could be explained by the weighting method.

HBW and HBO models had a Mcfadden $R^2$ value of 0.30 and 0.17, respectively, which is acceptable for the number of scenarios. For HBW, gender, employment, auto license, living area type and residence period in

**Table 3.** Value of Time across Income Groups, by Purpose

| Income group | HBW | HBO |
|---|---|---|
| <€1,500/month | 13.48 | 10.47 |
| >€1,500/month | 15.92 | 18.35 |

*Note:* HBO = home-based other; HBW = home-based work.

**Table 4.** Microscopic Transportation Orchestrator Value of Time across Income Groups, by Mode and Purpose

| Income group | Auto driver | | Auto passenger | | Transit | |
|---|---|---|---|---|---|---|
| | HBW | HBO | HBW | HBO | HBW | HBO |
| <€1,500 | 4.63 | 4.44 | 7.01 | 4.30 | 8.94 | 5.06 |
| €1,500–€5,600 | 8.94 | 6.11 | 13.56 | 8.31 | 17.30 | 9.78 |
| >€5,600 | 12.15 | 8.63 | 18.43 | 11.30 | 23.50 | 13.29 |

*Note:* HBO = home-based other; HBW = home-based work.

Munich were not statistically significant. The number of household autos and the interest in using RH were statistically significant and they increased the likelihood of selecting RH. RH use seemed to be more common with the younger age spectrum (18–39 years old). Bigger households and households located far from a transit station were also more likely to use RH services. For HBO, the model showed a similar behavior to the HBW model except that the distance to a transit station was not significant.

These results of RH being common with a younger group agree with the findings of (*10*) and (*13*). The higher use of RH services by households with fewer autos and all income groups supports the key findings in the TCRP report (*19*).

Based on the coefficients of time and cost, VOT for RH was derived. We distinguished two different income groups: <€1,500 and >€1,500. VOT was calculated by time and cost coefficients for RH using Equation 5 (*31*).

$$\text{Value of time (VOT)} = \frac{\beta\text{time}}{\beta\text{cost}} * 60 \left(\frac{€}{\text{hr}}\right) \qquad (5)$$

where

$\beta$time is the coefficient of time using RH; and

$\beta$cost is the coefficient of cost for an income group.

Table 3 presents the results for the estimated VOT across income groups for HBW and HBO trips.

Higher income groups being willing to use RH services more than the lower income groups for both trip purposes agrees with the findings of (*10*) in Toronto. The VOT for various modes, purposes and income groups used in MITO is presented in Table 4. For the lower income group, the VOT for HBW and HBO of RH is higher than transit and any other mode. In other words, lower income group users of other modes are likely to switch to RH. For the higher income group, the VOT for HBW and HBO of RH is between auto driver and auto passenger. This means that higher income group users would be willing to pay more for eliminating the task of driving.

## RH Service Scenarios

Figure 2 presents the comparison of mode split with and without ride-hailing (RH) in the base scenario (travel time as auto travel time, no waiting time and €3.0/km fare) at the following trip purposes: home-based work (HBW), home-based education (HBE), home-based shop (HBS), home-based other (HBO), non-home-based work (NHBW), and non-home-based other (NHBO).

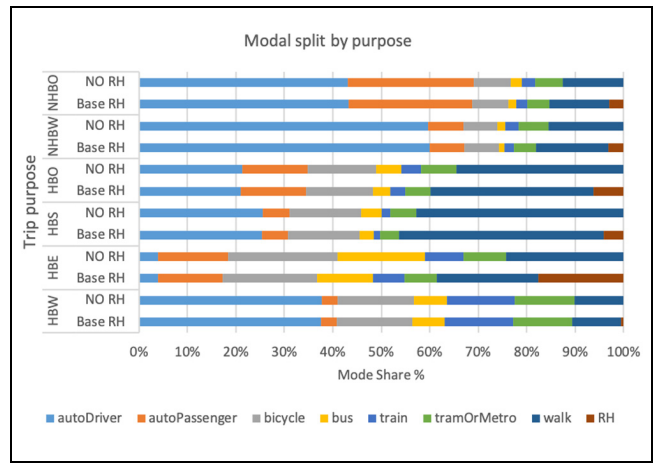

**Figure 2.** Modal split by purpose, with and without ride-hailing (RH), in the base scenario, showing home-based work (HBW), home-based education (HBE), home-based shop (HBS), home-based other (HBO), non-home-based work (NHBW), and non-home-based other (NHBO).

For HBW, HBS, HBO, NHBW and NHBO trips, the share for RH was less than 7% and a large proportion of the RH share was gained from public transit. The results agree with the survey model estimates, in which RH represented 8.04% of HBO trips and 4.66% of HBW trips. HBE trips had an RH share of around 18%, of which the majority share was gained from public transport (10%) and the active transportation modes (7%). These findings show RH trips mostly substituting transit services and active modes which is quite opposite to the RH effect in the Toronto study by (*10*) where RH complemented transit and active transport modes.

Figures 3–6 present the share of RH across various scenarios for the different trip purposes. In each figure, travel time is equal to one level (1, 1.1, 1.2 and 1.5) and travel cost varies. Figure 3 presents the RH share at TT = 1 or the least congested level and various costs. Compared with other trip purposes, HBW did not seem

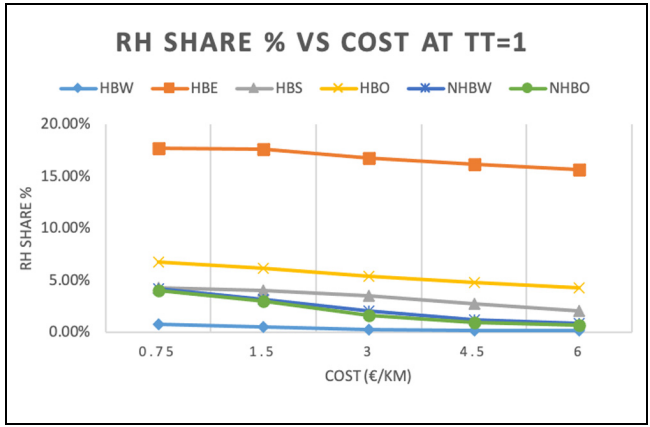

**Figure 3.** Ride-hailing (RH) share (%) versus cost (€/km) at travel time (TT) = 1, showing home-based work (HBW), home-based education (HBE), home-based shop (HBS), home-based other (HBO), non-home-based work (NHBW), and non-home-based other (NHBO).

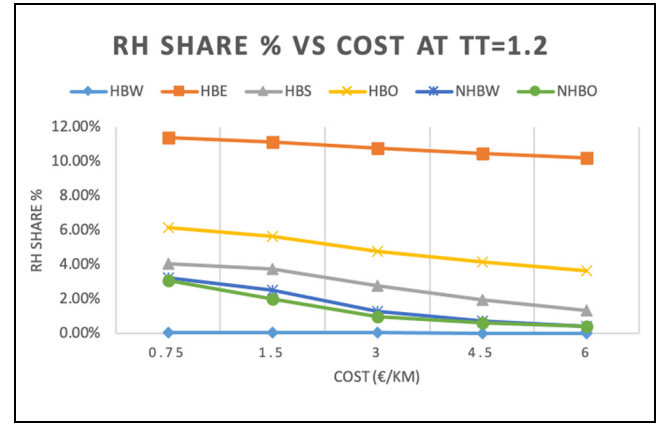

**Figure 5.** Ride-hailing (RH) share (%) versus cost (€/km) at travel time (TT) = 1.2, showing home-based work (HBW), home-based education (HBE), home-based shop (HBS), home-based other (HBO), non-home-based work (NHBW), and non-home-based other (NHBO).

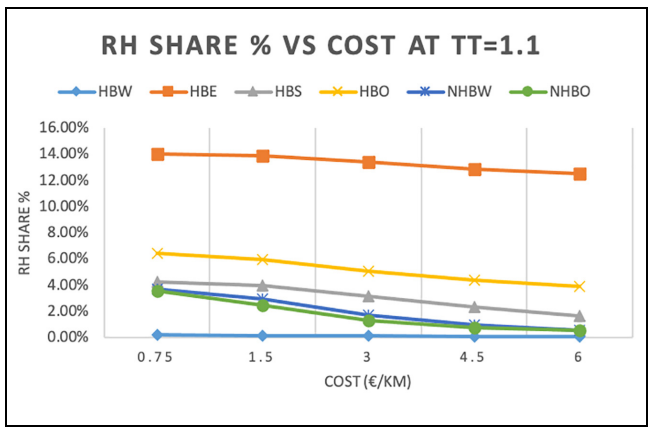

**Figure 4.** Ride-hailing (RH) share (%) versus cost (€/km) at travel time (TT) = 1.1, showing home-based work (HBW), home-based education (HBE), home-based shop (HBS), home-based other (HBO), non-home-based work (NHBW), and non-home-based other (NHBO).

to be very affected by the cost changes. The maximum variation (between lowest and highest cost) of HBW was less than 0.6% in comparison with 3% for other purposes. The trip purposes with highest sensitivity to cost were NHBW and NHBO.

At TT = 1.1, the share of RH (Figure 4) across all purposes was reduced compared with the base travel time scenarios, as expected. The variation of RH share with increasing cost was also reduced. Variation with cost follows the same pattern as in the base travel time scenario but with lower reduction. Similar conclusions may be obtained from the travel time increase by 20% (Figure 5).

The results for the highest travel time varied significantly from the previous scenarios (Figure 6). The share across all purposes was reduced, with the highest reduction being in the share of HBE, almost reduced to half compared with congestion at TT = 1.2 (from a maximum of 11.4% to a maximum of 6.4%). Further, this reduction was lower for other purposes: for HBO, the maximum varied from 6.2% to 4.6%, or for HBS from 4.1% to 3.6%. This indicates that HBE trips were more sensitive to travel time than other trip purposes. No HBW trips used RH when costs were higher than €0.75/km.

The largest share of RH trips for all trip purposes happened at the least congested level (TT = 1) and lowest cost level (€0.75/km) and the smallest share of RH trips happened at the most congested level (TT = 1.5) and highest cost level (€6/km). RH share for all purposes reflected a consistent behavior across scenarios when it came to changes in RH share with increasing cost. It is noticeable that the total RH trips share at TT = 1 and €3/km is equal to the RH share at TT = 1.1 and €1.5/km. There may be other combinations that balance costs and travel time, producing similar modal share for RH.

## Conclusion

The rapid growth and expansion of RH services with widespread claims about their benefits to other transportation modes, despite the lack of open data and limited research available on the topic, increases the importance of innovative research methodologies to gather representative data. Using Munich city as a case study, this project predicts the impact of RH on modal share based on the user's willingness to pay for such services. Despite a potential bias in responses, the results provided a preliminary understanding of the willingness to use RH services and the preferences for other similar transportation modes.

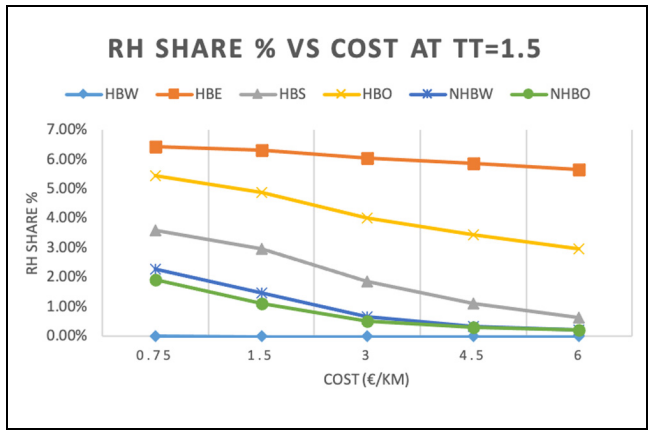

**Figure 6.** Ride-hailing (RH) share (%) versus cost (€/km) at travel time (TT) = 1.5, showing home-based work (HBW), home-based education (HBE), home-based shop (HBS), home-based other (HBO), non-home-based work (NHBW), and non-home-based other (NHBO).

Given that the number of survey responses was limited for estimating a mode choice to be applied to the entire study area and with all modes available in the choice set, we used the incremental logit approach to incorporate a new transport mode into an existing nested logit model calibrated for the study area. The survey responses were used for the assumptions on how RH utility may differ from the utility of the base mode, which was metro. Specifically, we derived the difference in travel time perception between RH and metro and the VOT for RH for two income categories. The results for the base scenario were similar to the survey responses, with estimated modal shares for RH of 8.04% for HBO and 4.66% for HBW.

Furthermore, we carried out a sensitivity analysis of RH modal share for different travel time and travel cost scenarios. The results indicate that RH ridership was more sensitive to travel time than travel cost: an increase from €1.5/km to €3.0/km reduced the ridership by 10%, but a 10% increase in travel time reduced the ridership by 13% for HBO and HBS trips. HBE trips were more sensitive to travel time than other trip purposes, and HBW trips were hardly ever made by RH (less than 1% in the most favorable scenario). This study can help the local region and policy makers understand the impacts of RH when making policy decisions and engineering developments. The share of RH trips over various congestion levels and costs allows decision makers to understand how demand behavior reacts against both variables.

There are some limitations with regards to the survey scenario designed, in which transit and auto scenarios remained constant and only one specific route was designed. The overrepresentation of some population segments was corrected by using weighted responses, although a higher representation of all population segments would be desirable. The VOT estimations were also produced with a simple MNL model in which some important attributes were not significant. Increasing the sample size from the survey, the representation of different population segments and the number of routes could improve the results. Further, the routes could be adapted to each respondent's commute trip and most common shopping or leisure trip, by origin, destination and time of day. For future work, the modeling of induced demand and choices between single occupancy RH vehicle and multiple occupancy (pooled ride) will be explored. Furthermore, once ride-hailing is more common in our study area, we will reconduct an RP survey with a redesign of the scenarios. This approach would better reflect the actual experience of RH users, similar to the study done by (*14*). The new survey will also be distributed differently to cover the entire population more homogeneously.

From the supply side, pickup and drop-off of passengers can be simulated from door-to-door to zone-to-zone or certain stops. Further, the results from the travel demand model for mode choice will be incorporated into a traffic assignment model to estimate the impact of RH on congestion levels and mobility in the city of Munich.

## Acknowledgment

We would like to acknowledge the valuable feedback and comments provided by Rolf Moeckel during the defense of the master thesis.

## Author Contributions

The authors confirm contribution to the paper as follows: study conception and design: M. Shoman and A. T. Moreno; data collection: M. Shoman; analysis and interpretation of results: M. Shoman; draft manuscript preparation: M. Shoman and A. Moreno. All authors reviewed the results and approved the final version of the manuscript.

## Declaration of Conflicting Interests

The author(s) declared no potential conflicts of interest with respect to the research, authorship, and/or publication of this article.

## Funding

The author(s) disclosed receipt of the following financial support for the research, authorship, and/or publication of this article: The research was completed with the support of the Technische Universität München Institute for Advanced Study, funded by the German Excellence Initiative and the European Union Seventh Framework Programme under grant agreement number 291763.